\documentclass[graybox]{svmult}

\usepackage{mathptmx}       
\usepackage{helvet}         
\usepackage{courier}        
\usepackage{type1cm}        
             \usepackage{subfigure}                  
\usepackage{makeidx}         
\usepackage{multicol}        
\usepackage[bottom]{footmisc}
\usepackage{url}

\usepackage[pdftex]{graphicx}

\makeindex             

\begin{document}

\title*{An application of Zero-One Inflated Beta regression models for predicting health insurance reimbursement}
\author{Fabio Baione, Davide Biancalana and Paolo De Angelis}
\institute{Fabio Baione, \at Department of Statistics, Sapienza University of Rome, Viale Regina Elena, 295 00161 Roma, Tel. +39 06 49255317, \email{fabio.baione@uniroma1.it}
\and Davide Biancalana, \at Department of Statistics, Sapienza University of Rome, Italy \email{davide.biancalana@uniroma1.it}
\and Paolo De Angelis, \at Department of methods and models for economics territory and finance, Sapienza University of Rome, Italy \email{paolo.deangelis@uniroma1.it}
}

%
%
\maketitle

\abstract{In actuarial practice the dependency between contract limitations (deductibles, copayments) and health care expenditures are measured by the application of the Monte Carlo simulation technique. We propose, for the same goal, an alternative approach based on Generalized Linear Model for Location, Scale and Shape (GAMLSS).
	We focus on the estimate of the ratio between the one-year reimbursement amount (after the effect of limitations) and the one year expenditure (before the effect of limitations).  We suggest a regressive model to investigate the relation between this response variable and a set of covariates, such as limitations and other rating factors related to health risk.
	In this way a dependency structure between reimbursement and limitations is provided.
	The density function of the ratio is a mixture distribution, indeed it can continuously assume values mass at 0 and 1, in addition to the probability density within $\left(0, 1\right) $. This random variable does not belong to the exponential family, then an ordinary Generalized Linear Model is not suitable. GAMLSS introduces a probability structure compliant with the density of the response variable, in particular zero-one inflated beta density is assumed. The latter is a mixture between a Bernoulli distribution and a Beta distribution.}

\section{Introduction}
\label{sec:1}
Insurance deductibles are a very important contract boundary which aim is to limit the abuse of reimbursement requests especially in health insurance policies. Generally, the data set of such contracts contains only censored and/or truncated values, but when the data contains the medical invoices, as well as the reimbursement values, it is possible to investigate the effect of deductibles in terms of the ratio of the truncated values and the overall loss, briefly referred as indicated deductible relativity (IDR). A standard reference for deductible pricing in actuarial science is in \cite{BrownLennox}. A practical approach to modeling the deductible rates in a ratemaking process is the adoption of regression models such as Generalized Linear Models (GLM), see \cite{Lee}, \cite{FreesLee}, \cite{FreesLee2} among others. Anyway, GLMs are used to predict aggregate claims of the truncated response variable using a log deductible covariate and their use is limited to the class of exponential family distributions. Our aim is to propose a different approach to measure the influence of deductibles on health care expenditures by focusing on a regression model, where IDR is the response variable. Anyway, the probability function of IDR in our framework belongs to a particular mixture distribution named Zero-One Inflated Beta (ZOIB) \cite{Ospina}. As a consequence we suggest to use a regression model of Generalized Linear Model for Location, Scale and Shape (GAMLSS) \cite{Rigby} type.

\section{Actuarial Framework} \label{subsec:2}

We consider an Health insurance company, whose portfolio is allocated on $H$-homogeneous sub-portfolios. Let:
\begin{itemize}
	\item $Y_i$ is the random variable (r.v.) expenditure for single episode, i.e. before the application of deductibles;
	\item $L_i$  be the r.v. reimbursement for single episode, i.e. after the application of deductibles;	
	\item $\textbf{x}_i$: be the row vector of the design matrix, providing information about policyholder or contract rating factors;
	\item $f$ the amount of deductible ;
	\item $M$ the amount of the out of pocket maximum.
\end{itemize}

Hence, we can state:
\begin{equation}
   \label{Lij}
   L_{i} = \min\left\lbrace  I_{\left( Y_i-f>0\right) }\cdot Y_i ;M\right\rbrace
\end{equation}
where $I_{\left( Y_i-f>0\right)}$ is the indicator r.v. for the event $Y_i-f>0$. Hence, if the invoice amount is higher than $f$ the health insurance pays the full expenditure within the limit fixed by $M$ and 0 otherwise.
In order to assess the influence of the deductible and the out of pocket maximum on the loss of the company, we focus on the random variable  $R_i=\frac{Y_i}{L_i}$, that is the proportion of expenditure reimbursed (Indicated Deductible Relativity IDR).\\
Letting $h$ index the number of the $H$ risk classes and  $\boldmath{D}_h$ be the $g_h$-elements set of policyholders belonging to $h$-th risk class, we can state  $\textbf{x}_{i \in \boldmath{D}_h}=\textbf{x}_h, \forall i \in D_h$.
Hence, in the following, $E(R|\textbf{x}_{i \in \boldmath{D}_h})=E(R|\textbf{x}_h)=E(R_h), \forall i \in \boldmath{D}_h $ is the mean conditioned to the $\textbf{x}_h$ vector of risk factors.\\
It is important to note that the r.v. $R_h$ can continuously assume values between zero and one, but $Prob\left( R_h=0\right) >0$ and $Prob\left( R_h=1\right) >0$. It means that the density function of $R_h$ is a mixture distribution, because $R_h$ is derived from a collection of two random variables:
the first one representing the event $(R_h=0 \cup R_h=1)$ and the second defining the level of
proportion conditioned to the event $\left( 0<R_h<1\right)$.
This mixture belongs to the so called Zero-One Inflated distributions as the density function of the IDR can continuously assume values mass at 0 and 1, in addition to the probability density within $\left(0, 1\right)$. To model the density between $\left(0, 1\right)$ we consider a Beta distribution. Then  the r.v. $R_h$ is assumed to be Zero-and-One Inflated Beta distributed.

We introduce a mixture between a Bernoulli distribution and a Beta distribution as proposed in (Ospina and Ferrari \cite{Ospina}). Specifically, assuming that the cumulative distribution function (hereafter cdf) of the generic random variable $R$ is:
\begin{equation}
   \label{J4}
   BEINF\left( r;p_o,p_1,a,b\right)  =\left( p_0 + p_1\right)\cdot Ber\left( r;\frac{p_1}{p_0+p_1}\right) +\left( 1-p_0 + p_1\right)\cdot Beta\left( r;a,b\right)
\end{equation}
where $Ber\left( \cdot;\frac{p_1}{p_0+p_1}\right)$ represents the cdf of a Bernoulli random variable with parameter $\frac{p_1}{p_0+p_1}$ and $Beta\left( r;a,b\right)$ is the Beta cdf, whose density function is:
\begin{equation}
   \label{beta_dens}
   beta(r;a,b)=\frac{\Gamma(a+b)}{\Gamma(a) \cdot \Gamma(b)}\cdot r^{a-1}\cdot (1-r)^{b-1}\   , \; \; \;  0<r<1
\end{equation}

We say that $R$ has a Zero-and-One Inflated Beta distribution (henceforth BEINF) (i.e $R \sim BEINF\left( p_o,p_1,a,b\right)$) with parameters $p_0$, $p_1$,
$a$ and $b$, if its density function (hereafter $beinf\left( r;p_o,p_1,a,b\right)$)  with respect to the measure generated by the mixture is given by:
\begin{equation}
   beinf\left( r;p_o,p_1,a,b\right) =\left\lbrace \begin{array}{cc}  p_0 & r=0\\
     p_1 & r=1\\
     \left(1-p_0-p_1\right)\cdot f(r;a,b) & \,\,\,\,\,\, otherwise
\end{array}
\right.
\end{equation}
It is worth noting that is $p_0=Prob\left( R=0\right) $,$p_1=Prob\left( R=1\right) $.

In order to consider a BEINF distribution for each $R_h$, we need to specify the dependence structure between the response variables and the set of covariates $\textbf{x}_h$ by means of a regression model.

\section{GAMLSS theory}
\label{sec:2}
Considering the features of the r.v.s previously introduced a Generalized Linear Model for Location, Scale and Shape (GAMLSS) seems to be an appropriate choice (see Rigby and Stasinopoulos \cite{Rigby} and \cite{Stas}) for our goal.
A GAMLSS model assumes independent observations $r_i$, $i=1, \dots, I$ from a random variable $R_i$, with probability density function $f\left( r_i|\boldmath{\theta^i}\right)$, conditional on $\boldmath{\theta^i}=\left( \theta_{1,i},\theta_{2,i},\theta_{3,i},\theta_{4,i}\right)=\left( \mu_i,\sigma_i,\nu_i,\tau_i\right)$
a vector of four distribution parameters, each of which can be a function to the explanatory variables. We shall refer to	$\left( \mu_i,\sigma_i,\nu_i,\tau_i\right)$ as the distribution parameters. The first two population distribution parameters $\mu_i$ and $\sigma_i$ are usually characterized as location and scale parameters, while the remaining parameter(s), if any, are characterized as shape parameters, e.g., skewness and kurtosis parameters, although the model may be applied more generally to the parameters of any population distribution, and can be generalized to more than four distribution parameters.
Rigby and Stasinopoulos define the original formulation of a GAMLSS model as follows. Let $\boldmath{r}^T=\left( r_1,\ldots,r_I\right) $ be the $I$ length vector of the response variable. Also for $k = 1, 2, 3, 4$, let $g_k(.)$ be known monotonic link functions relating the distribution parameters to explanatory variables by:
\begin{equation}
   \label{g_k}
   g_k\left( \boldmath{\theta}_k\right) =\eta_k=\boldmath{X}_k\beta_k +\sum_{j=1}^{J_k} Z_{jk}\gamma_{jk}, \,\,\,\,\, with \,\,\,\, k=1,2,3,4.
\end{equation}
where $\boldmath{\mu}$, $\boldmath{\sigma}$, $\boldmath{\nu}$, $\boldmath{\tau}$ and $\boldmath{\eta}_k$ are vectors of length $I$, ${\beta_k^T}=\left( \beta_{1k},\beta_{2k}, \ldots, \beta_{m_kk}\right)$ is a parameter vector of length $m_k$, $\boldmath{X}_k$ is a fixed known design matrix of order $I \times m_k $, $Z_{jk}$ is a fixed known $I \times q_{jk}$ design matrix and $\gamma_{jk}$ is a $q_{jx}$ dimensional random variable which is assumed to be distributed as $\gamma_{jk} \sim N_{q_{jk}}\left( \boldmath{0},\boldmath{G}_{jk}^{-1}\right)$, where $\boldmath{G}_{jk}^{-1}$ is the (generalized) inverse of a $q_{jk} \times q_{jk}' $ symmetric matrix $\boldmath{G}_{jk}=\boldmath{G}_{jk}\left( \boldmath{\lambda}_{jk}\right)$, which may depend on a vector of hyperparameters $\boldmath{\lambda}_{jk}$, and where if $\boldmath{G}_{jk}$ is singular then $\gamma_{jk}$ is understood to have an improper prior density function proportional to $exp\left( -\frac{1}{2}\gamma_{jk}^{T}\boldmath{G}_{jk}\gamma_{jk}\right)$.

The population probability (density) function $f\left( y|\boldmath{\theta}\right)$ in model (\ref{g_k}) is deliberately left general with no explicit conditional distributional form for the response variable. As in the previous section we have introduced a density function particularly useful for our ratemaking purposes, the Zero-One Inflated Beta, then we can express the following parametrization as proposed in \cite{Rpack}:
\begin{equation}
   \label{mu}
   \boldmath{\mu}=\frac{a}{a+b};         \;\;\; \boldmath{\sigma}=\frac{1}{a+b+1};   \;\;\;
   \boldmath{\nu}=\frac{p_0}{1-p_0-p_1}; \;\;\; \boldmath{\tau}=\frac{p_1}{1-p_0-p_1}
\end{equation}

\section{Some numerical Results}

We set our model on a database from an Italian health insurance company between years 2017 and 2018, the company offers two kind of covers: the first for  surgery and the second for diagnostic. For this two products, we observe a number of policyholder exposed $I_s=63,790$ and $I_d=58,994$ respectively. We apply a GAMLSS on the observed IDR as response variable, where the covariates are a class of deductibles with 3 levels (1 lowest).

A logit function is used for $\mu$ and $\sigma$, as these parameters must be included between 0 and 1 (see \ref{g_k}). For $\nu$ and $\tau$ we choose a logarithmic function, since the lowest value of the AIC is obtained. Random effect are not considered in this study.\\
Hence, equations (\ref{J1}), (\ref{J2}), (\ref{J3}) and (\ref{J4}) reports the model assumed for the four parameters
\begin{eqnarray}
   logit\left(\boldmath{\mu}\right)    &=& \beta_{1,0}+deductible_2\cdot\beta_{1,1}
                                           +deductible_3\cdot \beta_{1,2} \label{J1} \\
   logit\left(\boldmath{\sigma}\right) &=& \beta_{2,0} \label{J2} \\
   log\left( \boldmath{\nu}\right)     &=& \beta_{3,0}+deductible_2\cdot\beta_{3,1}
                                           +deductible_3\cdot \beta_{3,2} \label{J3} \\
   log\left( \boldmath{\tau}\right)    &=& \beta_{4,0}+deductible_2\cdot \beta_{4,1}
                                           +deductible_3\cdot \beta_{4,2} \label{J4}
\end{eqnarray}

\begin{figure}[!ht]
	\sidecaption[t]
		\centering
	\subfigure[$p_0$: surgery]{
		\includegraphics[ width=0.45\textwidth]{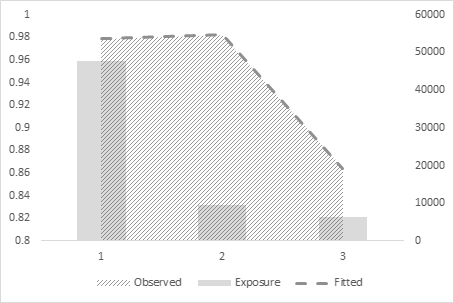}}
	\centering
	\subfigure[$p_0$: diagnostic]{
		\includegraphics[ width=0.45\textwidth]{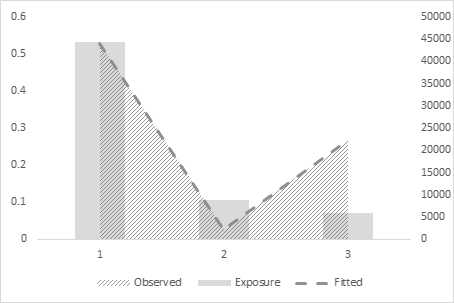}}
	\centering
		\subfigure[$p_1$: surgery]{
		\includegraphics[ width=0.45\textwidth]{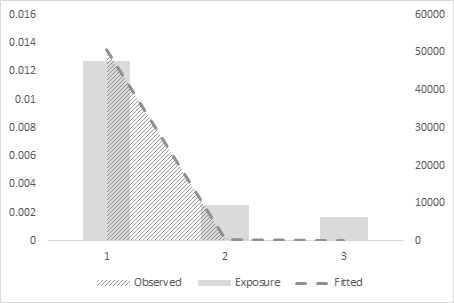}}
	\centering
	\subfigure[$p_1$: diagnostic]{
		\includegraphics[ width=0.45\textwidth]{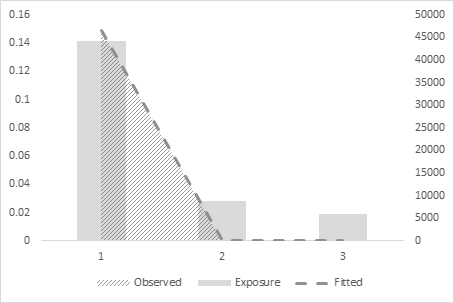}}
	\centering
	\subfigure[$\mu$: surgery]{
		\includegraphics[ width=0.45\textwidth]{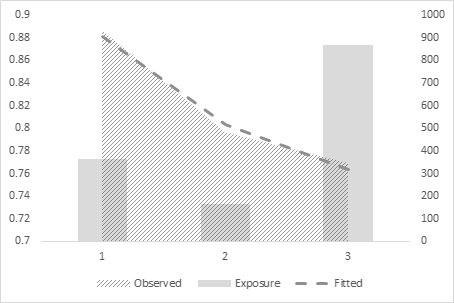}}
	\centering
\mpicplace{25pt}{20pt}
	\subfigure[$\mu$: diagnostic]{
		\includegraphics[ width=0.45\textwidth]{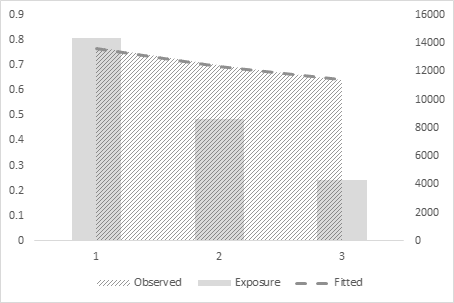}}
	\caption{Goodness of fit of GAMLSS with BEINF distribution.}
		\label{fitting}
\end{figure}

In Table \ref{tab: estimate} and Figure \ref{fitting} the fitted values are reported and compared with the observed values.
\begin{table}[!ht]
	\caption{Observed vs Fitted values.}
	\begin{center}
		\small
		\begin{tabular}{c|c|ccc|ccc|ccc}
			\hline\noalign{\smallskip}
   Branch & Deductible&\multicolumn{3}{c}{Beta}&\multicolumn{3}{c}{$p_0$}&\multicolumn{3}{c}{$p_1$}\\\cline{3-11}
   	
& &Exposure& Observed&Fitted&Exposure&Observed&Fitted&Exposure&Observed&Fitted\\
	\hline\noalign{\smallskip}
Surgery&Level 1&                            366 &88.72\%&88.08\%&     47,845 &97.88\%&97.88\%&     47,845 &1.35\%&1.35\%\\
&Level 2&                            167 &79.66\%&80.30\%&        9,566 &98.24\%&98.24\%&        9,566 &0.01\%&0.01\%\\
&Level 3&                            867 &76.90\%&76.37\%&        6,379 &86.41\%&86.41\%&        6,379 &0.00\%&0.00\%\\
\hline\noalign{\smallskip}
Diagnostic&Level 1&                      14,358 &77.04\%&76.58\%&     44,245 &52.67\%&52.67\%&     44,245 &14.88\%&14.88\%\\
&Level 2&                        8,649 &70.08\%&69.47\%&        8,889 &2.70\%&2.70\%&        8,889 &0.00\%&0.01\%\\
&Level 3&                        4,298 &64.10\%&64.12\%&        5,860 &26.66\%&26.66\%&        5,860 &0.00\%&0.00\%\\

			\hline\noalign{\smallskip}
		\end{tabular}
	\end{center}
	\label{tab: estimate}
\end{table}

As observable, the GAMLSS shows a very good quality of fitting as the probability of a 0 and 1 are perfectly matched whereas the values in $(0,1)$ have a distance of less then 1\% in each branch.

\end{document}